\documentclass[acmtog]{acmart}

\usepackage{caption}
\usepackage{subcaption}
\usepackage{booktabs}

\usepackage{booktabs} 
\usepackage{balance}
\usepackage{enumitem}
\usepackage[colorinlistoftodos]{todonotes}
\usepackage{varwidth}
\usepackage[dvipsnames]{xcolor}
\usepackage{cuted}
\usepackage{tabularx}
\usepackage{subcaption}
\usepackage{absfigure}

\usepackage[ruled]{algorithm2e} 

\SetAlFnt{\small}
\SetAlCapFnt{\small}
\SetAlCapNameFnt{\small}
\SetAlCapHSkip{0pt}

\makeatletter
\g@addto@macro\normalsize{%
  \setlength{\abovedisplayskip}{4pt}%
  \setlength{\belowdisplayskip}{4pt}%
  \setlength{\abovedisplayshortskip}{4pt}%
  \setlength{\belowdisplayshortskip}{4pt}%
}
\makeatother

\acmJournal{TOG}

\copyrightyear{2026}
\acmYear{2026}
\setcopyright{cc}
\setcctype{by}
\acmConference[SIGGRAPH Conference Papers '26]{Special Interest Group on Computer Graphics and Interactive Techniques Conference Conference Papers}{July 19--23, 2026}{Los Angeles, CA, USA}
\acmBooktitle{Special Interest Group on Computer Graphics and Interactive Techniques Conference Conference Papers (SIGGRAPH Conference Papers '26), July 19--23, 2026, Los Angeles, CA, USA}
\acmDOI{10.1145/3799902.3811213}
\acmISBN{979-8-4007-2554-8/2026/07}

\begin{document}
\title[Retrofitting Existing 3D Objects with Surface-Conforming Capacitive Sensing]{Retrofitting Existing 3D Objects with \\ Surface-Conforming Capacitive Sensing}

\author{Andela Ilic, 
Junpeng Gao, 
Zhipeng Li, 
Yijing Jiang, 
Rachel Schuchert, 
Manuel Meier, 
Philipp Herholz, 
and Christian Holz}

\affiliation{
  \institution{Department of Computer Science, ETH Zürich}
  \city{Zürich}
  \country{Switzerland}
}

\renewcommand\shortauthors{Ilic et al.}

\begin{teaserfigure}
  \centering
  \includegraphics[width=\textwidth]{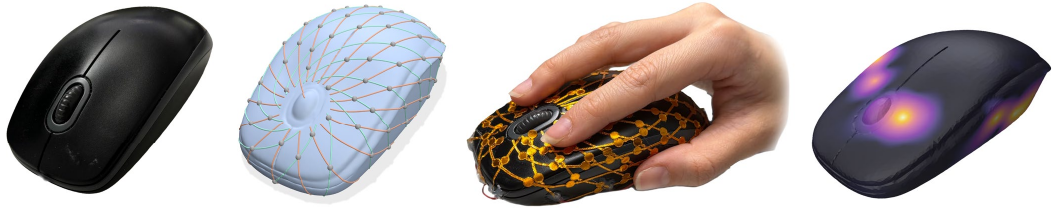}%
  \vspace{-2mm}%
  \caption{%
    We propose a computational fabrication pipeline that equips existing 3D objects with mutual-capacitance sensing for touch input.
    Starting from a 3D scan of a physical object~(left), our method generates two layers of electrode curves that conform to the object's surface geometry, tailoring their layout with linear optimization for sensor coverage and distribution~(center).
    We fabricate these curves from copper foil using a vinyl cutter, attach them to the surface, and connect them to a mutual-capacitance controller for live sensing and visualization~(right).%
  }%
  \label{fig:teaser}%
  \vspace{2mm}%
\end{teaserfigure}

\begin{abstract}
Augmenting the surface of 3D objects with capacitive sensing is challenging when their volumes cannot be modified.
In this paper, we present a generative computational fabrication pipeline that retrofits surface-only sensor layouts to 3D geometries for multi-touch interaction. 
Our method scans a real-world object to obtain its 3D mesh, generates and optimizes a 3D sensor design of drive and sense lines for mutual-capacitance sensing under physical and hardware constraints, and unfolds the design into individual 2D stencils that can be cut from conductive material.
Our fabrication pipeline cuts these stencils from thin copper foil with a vinyl cutter and then assists manual sensor attachment by projecting the sensor design onto the dynamically registered real-world object.
We connect the resulting electrode mesh to a mutual-capacitance scanning controller and resolve touch interaction in real time.
We demonstrate our approach with four 3D geometries and evaluate our method and fabrication pipeline on them.
\end{abstract}



\maketitle
\section{Introduction}

Touch-sensitive surfaces enable direct interaction with physical objects and expand the ways we engage with digital systems.
A wide range of domains benefit from this functionality, including toys, tangible interfaces, and, not least, interactive devices.
On them, touch sensing registers both contact events and their spatial locations~\cite{grosse-puppendahl_finding_2017}, which becomes increasingly powerful when entire surfaces act as continuous sensors for flat~\cite{zhang_wall_2018,wessely2020sprayable,parilusyan_sensurfaces_2022,rus_recognition_2014} and non-planar~\cite{palma2024capacitive,nittala_multi-touch_2018,savage_midas_2012,schmitz_capricate_2015,villar2009mouse20} geometries.

\begin{figure*}[t]
    \centering
    \vspace{-2mm}%
    \includegraphics{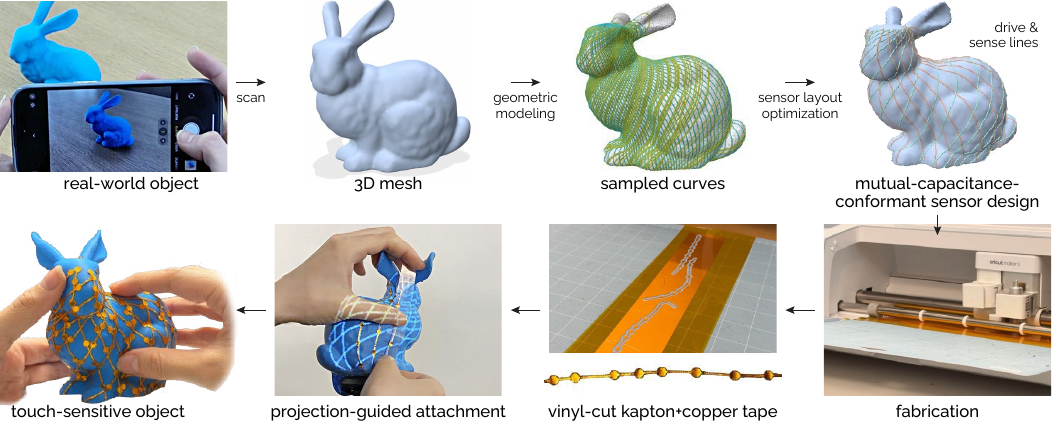}%
    \caption{Our computational fabrication pipeline comprises real-world object scanning, 3D sensor layout generation, optimization for hardware constraints and sensor surface coverage, conductor fabrication, projection-guided sensor attachment, and interactive touch sensing and visualization.}%
    \vspace{-1mm}%
    \label{fig:pipeline}%
\end{figure*}

However, covering the surface of a given 3D object with a continuous touch sensor is technically difficult.
This has often required fabricating the object with embedded sensing components, either through 3D printing conductive materials~\cite{palma2024capacitive,schmitz_capricate_2015} or embedding discrete elements, such as pressure sensing~\cite{parzer2017smartsleeve} or fiber-based light sensing~\cite{willis2012printed}.
These methods typically trade off spatial resolution and practicality.
Interesting exceptions to this are commercial devices such as the Apple Magic Mouse or Microsoft's TouchMouse~\cite{villar2009mouse20}; they integrate mutual-capacitance for curved surfaces, albeit for moderate curvature and simple geometry that could accommodate a handcrafted sensor layout.

To augment \emph{existing objects} whose volume cannot be modified, reproduced, or redesigned, previous approaches have instrumented select sub-regions with sensors~\cite{savage_midas_2012,kawahara2013instant,pourjafarian_multi-touch_2019}.
Alternatives have relied on external sensing through cameras to monitor hand-object interaction from one~\cite{cao_reconstructing_2021,izadi_kinectfusion_2011} or more vantage points~\cite{brahmbhatt_contactpose_2020}.
Common in such setups is occlusion during direct manipulation, such that estimating touch contact from vision alone requires resolving both the occlusion of fingers~\cite{streli_touchinsight_2024} and ambiguous object surface geometry~\cite{fan_hold_2024}.

In this paper, we present a generative computational fabrication pipeline for augmenting 3D geometries with capacitive sensing for multi-touch interaction.
Our approach generates surface-only sensor layouts for an object's 3D geometry and requires no modification of its volume to establish real-time multi-touch input across its surface.

\vspace{-2mm}
\subsection{Mutual-capacitance sensing across a 3D surface}

Fig.~\ref{fig:pipeline} shows our fabrication pipeline for generating a mutual-capaci\-tance sensor layout that conforms to an existing real-world object.

We first scan the object to obtain its 3D mesh for geometry modeling.
Our method then generates two dense layers of smooth curves along the surface, one for drive lines and one for sense lines in a mutual-capacitance sensing configuration, such that only curves from different layers intersect.
Intersections between layers define the spatial distribution of input locations on the object surface.

Next, we sample curves from both layers to uniformly distribute sensing intersections across the surface under geometric, hardware, and fabrication constraints.
We cast this as an optimization problem that maximizes surface coverage and intersection uniformity.

Finally, our fabrication pipeline unfolds the generated 3D curves onto the 2D plane and exports the resulting vector outlines as individual stencils, cut from conductive material using a vinyl cutter.
To assist manual attachment, our system projects the optimized sensor design directly onto the 3D object using dynamic registration via an RGB-D camera.
After attaching the electrodes to the target object's surface and connecting them to a digitizer, our signal processing pipeline resolves touches and contact areas across the 3D surface in real time and maps them onto the 3D geometry for visualization.

Our target use cases include objects that cannot be refabricated or internally modified (e.g., a \emph{functional} mouse, ceramic figurine, or wooden toy), as well as thin objects (e.g., plates) without space for interior routing.
Our system supports standard touch interactions---tap, swipe, pinch, and multi-touch---on 3D surfaces, such as sliding along curved handles or tapping objects for region-specific input.

\subsection{Contributions}

We contribute a computational method that generates and fabricates surface-only mutual-capacitance sensor layouts for existing 3D objects while preserving their interior and structure, including:
\begin{enumerate}[leftmargin=*,topsep=2pt,labelsep=4pt]
\item a geometric method for generating surface-intrinsic multi-touch sensor layouts that retrofit existing 3D objects,

\item a constrained optimization method for determining fabrication-ready sensor layouts with broad and uniform coverage under mutual-capacitance, hardware, and fabrication constraints, and

\item a fabrication and assembly process that unfolds the generated 3D routings into planar conductive traces and guides their attachment to the physical object using dynamic projection.
\end{enumerate}

\vspace{-2mm}%
\section{Related work}

\subsection{Mutual-capacitance touch sensing}

\begin{figure}[t]
    \centering
    \includegraphics[width=\linewidth]{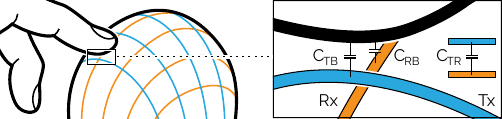}%
    \vspace{-2mm}%
    \caption{Mutual-capacitance sensing model with drive line~(Tx), sense line~(Rx), finger~(B), and resulting capacitors.
    Capacitors to GND and the finger's RC circuit to GND are omitted for simplicity.}%
    \label{fig:mutual_capacitance}%
    \vspace{-4mm}%
\end{figure}

Widely adopted, mutual-capacitance sensing detects touch by measuring the change in capacitive coupling between two sets of electrodes~\cite{zimmerman_applying_1995}.
Drive~(Tx) and sense~(Rx) lines form orthogonal grids in which electrode pads at each intersection establish capacitors (Fig.~\ref{fig:mutual_capacitance}).
A scanning controller sequentially drives Tx lines while measuring responses on Rx lines to create a 2D map of coupling changes caused by finger proximity.
This localizes simultaneous touches and has made capacitive sensing the default in phones, tablets, etc. (see the survey by \citet{grosse-puppendahl_finding_2017} for capacitive sensing designs and applications).

The sensor layout for mutual-capacitance measurements imposes strict geometric constraints:
conductors must not intersect within the same layer, and each drive line must intersect each sense line at most once to avoid spatial ambiguity in localizing measured changes in mutual capacitance.
Designing sensors for non-planar 3D surfaces thus introduces challenges in routing, intersection control, and fabrication.
The Multitouch-Kit proposed by \citet{pourjafarian_multi-touch_2019} showed a promising direction and achieved finger-scale resolution using dense Tx/Rx grids that can be conformed to mildly curved surfaces.
Related to our fabrication pipeline, Multitouch-Kit assumes that the sensing layout remains a developable 2D pattern: as surface curvature increases or topology becomes complex, maintaining electrode alignment and spatial correspondence becomes a nontrivial developability problem that is not addressed by Multitouch-Kit.\looseness=-1

\subsection{Touch sensing on 3D objects}

Previous work has explored a variety of techniques to augment 3D objects with touch sensing.
Examples of capacitive sensing include embedding conductive traces during 3D printing and manufacturing~\cite{10.1145/2910674.2910690, burstyn2015printput}, post-processing 3D surfaces using screen printing~\cite{olberding2014printscreen} or inkjet printing~\cite{kawahara2013instant, khan2019soft}, and augmenting objects with dedicated electrodes~\cite{zhang2019postureaware}. 
Spray coating with conductive materials~\cite{zhang2018pulp, wessely2020sprayable}, hydrographic transfers~\cite{groeger2018objectskin}, and functional adhesive stickers or patches~\cite{klamka2020rapid, strohmeier2018zpatch} can also augment existing objects with capacitive sensing.
Therefore, while prior approaches implement sparse or single-electrode sensing and require per-object calibration to localize touches, our grid-based approach inherently implements \emph{spatial} capacitive sensing for multi-touch localization.

Closely related is electric field sensing (EFS).
ElecTrick applies conductive paint to an object for large-area touch sensitivity~\cite{zhang2017electrick}.
Touch is localized via tomographic reconstruction between \emph{boundary-attached electrodes} and per-object \emph{calibration}.
This means that position estimates are constrained to regions spanned between electrodes and depend on a learned inverse mapping.
It is unclear, however, how well EFS can scale spatial sensing to general 3D surfaces or closed geometries where no clear electrode boundary exists and geometric correspondence to the surface is required.

Beyond capacitive sensing, resistive sensing has been applied to curved and soft surfaces to detect touch and pressure input.
Example materials to coat existing curved objects include stitched conductive threads~\cite{villar2018project}, zebra tiles~\cite{parzer2017smartsleeve}, or deformable substrates~\cite{sundholm2014smart}.
Alternatively, optical approaches have enabled touch sensing on curved surfaces by custom-shaping acrylic into optical waveguides~\cite{roudaut2011touchcurved}.\looseness=-1

Recent fabrication-oriented approaches have explored routing with low complexity for input detection.
\citet{sakurakakehi} demonstrated a single-stroke conductive-path touch sensor for 3D-printed objects that minimizes path complexity.
\citet{Bae_2025} proposed a single-wire interior approach for touch sensing on 3D objects, which reduces instrumentation overhead but constrains spatial resolution to the wire's trajectory.

\subsection{Geodesic layouts}

Vector field design has been widely used to guide the sampling of curves on surfaces.
Designing tangent vector fields with user control is well-studied~\cite{10.1145/2897826.2927303, 10.1145/3084873.3084921} and has been applied to generate field-aligned, fabrication-aware planar polylines on surfaces~\cite{10.1145/2537852}.
Given a set of user-defined constraints, previous methods have used edge-based representations to construct tangent vector fields~\cite{10.1145/1276377.1276447, Crane:2010:TCD}.
Others used vertex representations~\cite{10.1145/2870629} where users provide example fields or draw strokes directly on the surface.
Beyond vector fields, several recent projects directly sample curves on object surfaces~\cite{10.1145/3658158,10.1145/3439429}.

From a geometry-processing perspective, our problem is related to tubular decomposition and curve-network simplification. 
Shape Representation by \citet{Schueller:Zippables:2018} addresses decomposition into topological cylinders via tubular parameterization. 
Similarly, Stability-Aware Simplification of Curve Networks~\cite{10.1145/3528233.3530711} is relevant to our pruning stage, which balances structural constraints and coverage objectives under optimization.
We refer the reader to the survey by \citet{inversion-free} for further information on mapping-based alternatives and their robustness trade-offs.

For the \emph{spatial} mutual-capacitance sensing, curves sampled on the surface of 3D objects must satisfy specific constraints as mentioned above.
\citet{palma2024capacitive} presented a fabrication method that embeds electrodes on the surfaces of 3D-printed objects, requiring electrode wiring to be routed through hollow tubes inside the 3D print.
Their approach partitions surfaces into quadrilateral patches for the traditional grid-based input localization, but it requires access to and modification of the object's interior.
Even before, \citet{10.1145/2910674.2910690} fabricated Tx and Rx wires directly within 3D-printed objects during fabrication and similarly relied on internal routing and fabrication-time layout decisions.
In contrast, we address the setting of deploying mutual-capacitance touch sensing on existing objects through \emph{surface-only} instrumentation.

\section{Method overview}

Our approach models surface-based mutual-capacitance sensing on a 3D object as a geometric curve-routing problem on its surface.
Our formulation treats each conductor as an independent surface-intrinsic curve, with sensing locations defined by pairwise intersections between curves from two layers.
This abstraction enables spatial touch sensing on 3D geometries.

\vspace{-2mm}%
\subsection{Problem definition}
\label{sec:problem_definition}

We define two sets of surface curves, $A$ for drive lines and $B$ for sense lines, subject to the following constraints:

\begin{enumerate}[leftmargin=*,topsep=3pt]
    \item Curves within the same layer must not intersect to ensure proper and unambiguous drive and sense behavior.

    \item A curve in one layer may intersect a curve in the other layer at most once, so that each measurement corresponds to a unique sensing location.

    \item All curves must connect to the object's base, which accommodates electronic connections to the scanning controller.
\end{enumerate}

\noindent
We formalize these constraints in Section~\ref{sec:optimization} and extend them with fabrication feasibility constraints.

\section{Scanning real-world objects}

Our pipeline begins by acquiring the 3D mesh of a given real-world object.
Our system integrates SAM3D~\cite{yao2025sam3d3dobject} to continuously track a 3D object moved and rotated in front of an Azure Kinect camera to extract its initial surface geometry.
We load the scan into MeshLab for mesh repair, downsampling, and isotropic remeshing to standardize the input into our geometry pipeline.

\section{Geometric modeling to sample intrinsic curves}

Given a 3D mesh, regions of the surface are designated for touch sensing, along with a non-sensing region that accommodates wiring to the scanning controller. Fig.~\ref{fig:curve_sample} illustrates this process.
In practice, this non-sensing region corresponds to the object base where it rests on a supporting surface.
Starting from a seed face, we identify the contiguous base region by expanding across neighboring faces whose normals deviate by less than an angular threshold, thereby detecting the perimeter of the base.
For objects such as cups or the bunny, this region typically spans the full footprint of the object.

On the remaining sensing surface, two layers of smooth, surface-intrinsic curves are generated to represent drive (Tx) and sense (Rx) conductors.
Curve generation is guided by discrete trivial connections~\cite{Crane:2010:TCD}, which define consistent transport of tangent directions across mesh edges.
This intrinsic formulation enables coherent curve families on arbitrary meshes without requiring a global surface parameterization, while ensuring that all conductor paths remain confined to the surface.
Explicit intrinsic tracing is preferred over parameterization because it avoids mesh cutting and surgery required by parameterization, and it naturally enforces the single-intersection constraint without post-processing.
The overall layout of each curve family is controlled by a small set of singularities that determine the topology of the induced direction fields.\looseness=-1

\subsection{Computing trivial connections }

We follow~\citet{Crane:2010:TCD} and represent a discrete trivial connection by assigning each mesh edge a rotation angle that specifies how a tangent vector is transported between adjacent triangles.
The vector field assigns one oriented vector per face, with the trivial connection represented as an angle per dual edge.
Given a connection, we trace curves by integrating within a triangle until reaching an edge, transporting the tangent direction using the stored rotation, and continuing in the neighboring triangle.

All examples in this paper use genus~0 meshes and singularities of index~1.
Singularity placement is defined relative to the non-sensing base region.
We grow the first exclusion region by face expansion until either a geodesic-distance threshold or a normal-deviation threshold is reached, and we define the second exclusion region using geodesic distance alone. 
The Tx and Rx layers use different singularity placements, as described next.
(See~\citet{Crane:2010:TCD} for discretization details and mathematical properties.)

\subsection{Intrinsic curve sampling}
\label{sec: intrinsic_curve_sampling}

Given a trivial connection and its associated singularities, curves are sampled by tracing integral curves on the surface from prescribed seed locations and initial directions.
Tracing proceeds until a termination condition is met.

The Tx layer places a singularity at the center of the non-sensing base region, and curves are traced from a seed point defined as the point on the sensing surface with maximum geodesic distance from the base region.
Initial directions are sampled in the tangent plane, yielding curve families that distribute evenly across the sensing surface.
The Rx layer places two adjacent singularities at the center of the non-sensing base region, and curves are traced from a set of starting locations along the base boundary.
Initial directions are sampled relative to the boundary normal, producing curves that extend from the base into the sensing region, promote intersections with the Tx layer, and remain connected to the base.

Curve tracing terminates as trajectories approach singularities.
Tracing can also halt earlier when vectors on adjacent faces both point toward their shared edge. We detect this case explicitly; it occurs rarely in our pipeline, in about 3\% of traces, due to isotropic remeshing preprocessing. Intra-layer intersections are additionally prevented by constraint C1 in the integer linear programming formulation.
In addition to reflecting the topology induced by the connection, this criterion improves fabrication robustness, as curves tend to exhibit increased curvature near singularities.

\begin{figure*}[t]
    \centering
    \includegraphics[width=\linewidth]{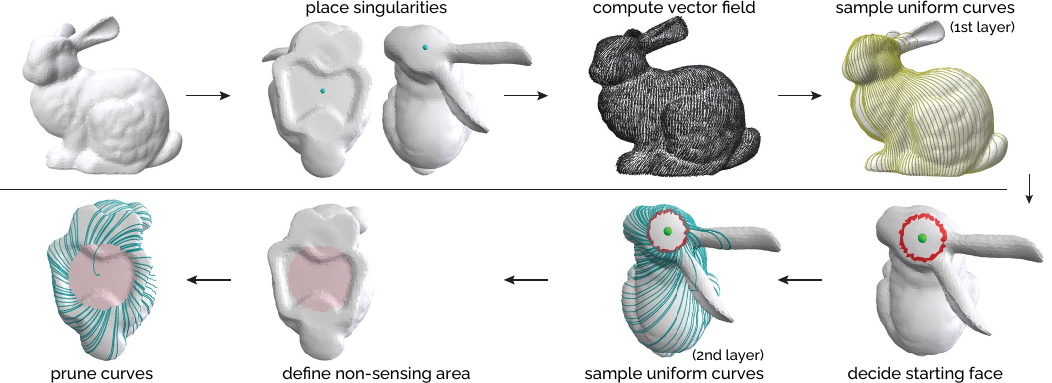}%
    \vspace{-3mm}%
    \caption{The curve layout is sampled as follows. For a 3D mesh, we first place singularities in non-sensing regions of the surface (e.g., the bunny’s base and head). Next, we compute a vector field and uniformly sample the first layer of curves. We then select starting faces for the second layer and sample curves at different angles. Finally, we refine the non-sensing regions and trim the curves at their boundaries.}
    \label{fig:curve_sample}
\end{figure*}

\subsection{Implementation}

We sample the two layers of intrinsic curves using two different strategies to ensure complementary coverage of the surface (Fig.~\ref{fig:curve_sample}).
For the first layer, we generate curves by sampling initial directions at fixed angular intervals in the tangent plane. 
All curves originate from one singularity and extend toward the other singularity. 
These curves are sampled at uniform angular intervals, resulting in a regularly spaced and approximately uniform layout.
For the second layer, we vary the curve orientation to complement the first layer.
We first select starting faces whose geodesic distance to one singularity is identical. 
Starting faces are selected by increasing the geodesic-distance threshold around the singularity until we obtain the required number of candidate faces for tracing.
From each starting face, curves are traced toward the opposite singularity with an initial angular offset. 
We sample this initial angle in the range \([-50^\circ, 50^\circ]\) at \(10^\circ\) intervals, generating one candidate curve set per angle.
A wider angular range produces undesired multiple intersections between curve pairs, while finer angles increase runtime without improved quality.
We sample 100~curves for each layer.

\section{Design optimization for capacitive sensing}
\label{sec:optimization}

Having sampled two layers of intrinsic curves on the object surface, we must now ensure that they meet the requirements for mutual-capacitance sensor designs.
Our method can sample large numbers of smooth curves, but fabricating them for real-world objects dictates additional feasibility constraints.
Beyond the constraints listed above, these constraints include augmenting Tx and Rx line intersections with small electrode pads to increase the signal-to-noise ratio for sensing touches, which may intersect with other curves, and accounting for the digitizer's limited number of Tx and Rx pins.

Therefore, we adapt the layout of Tx and Rx lines by selecting the optimal subset of curves from each layer such that the intersections between the two layers are distributed as evenly as possible across the surface.
A second optimization stage then selects a subset of intersections at which pads are placed while maintaining sufficient spatial separation between them.
We formulate this layout selection as an integer linear programming problem.

\subsection{Input \& Output}

The sensor design optimization proceeds in two phases. 
The input consists of two layers of sampled intrinsic curves.
Let the first and second layers be denoted as
\begin{equation}
    A = \{a_1, a_2, \ldots, a_m\}, \quad 
    B = \{b_1, b_2, \ldots, b_n\},
\end{equation}
where each curve \( a_i \in A \) and \( b_j \in B \) is an intrinsic curve sampled using the method described in \autoref{sec: intrinsic_curve_sampling}.

\paragraph{Phase~1: Curve selection.}
We select subsets \( A' \subseteq A \) and \( B' \subseteq B \) such that the intersections between curves in \( A' \) and \( B' \) are distributed evenly over the object surface.

\paragraph{Phase~2: Electrode pad placement.}
We maximize the number of electrode pads placed at intersections while enforcing a minimum distance constraint between any pair of selected intersections to avoid signal interference.
From the intersections between the curve subsets \( A' \) and \( B' \), we select a subset at which electrodes are placed. 

\paragraph{Optimization output}
is
(1)~the selected subsets of curves \( A' \) and \( B' \), and
(2)~the subset of intersections at which electrode pads are placed for each selected curve.

\subsection{Phase 1: Curve selection details}

Let
\vspace{-1.625em}
\begin{equation}
    I = \{\, i_{(a_i, b_j)} \mid a_i \in A,\; b_j \in B \,\},
\end{equation}
denote the set of all candidate intersections, where \( i_{(a_i, b_j)} \) represents the intersection point between curves \( a_i \) and \( b_j \), if it exists.
To quantify surface coverage, we uniformly sample the object surface to obtain a reference set
\begin{equation}
    S = \{ s_1, s_2, \ldots, s_K \},
\end{equation}
which serves as a discrete approximation of the surface geometry during optimization.
For each intersection \( i \in I \), we compute its geodesic distance to the closest sampled surface point:
\begin{equation}
    d_i = \min_{s_k \in S} \, \mathrm{dist}_{\mathcal{M}}(i, s_k).
\end{equation}
\( \mathrm{dist}_{\mathcal{M}}(\cdot,\cdot) \) denotes the geodesic distance along the surface manifold. 
We select subsets of curves from the two layers such that the subset of active intersections \( I' \subseteq I \) minimizes the maximum distance between any active intersection and its closest sampled point:
\begin{equation}
    \textstyle
    \min \; d_{\max}, \quad
    \text{where } d_{\max} = \max_{i \in I'} d_i .
\end{equation}
This objective ensures that every region of the surface lies within a bounded distance of at least one sensing location.

As a secondary objective, we minimize the average distance
\begin{equation}
    \textstyle
    \min \; \frac{1}{|I'|} \sum_{i \in I'} d_i,
\label{eq:eq6}
\end{equation}
to discourage clusters and promote even spatial distribution.

We specify several constraints required by grid-based mutual-capacitance and the hardware limitations.

\emph{C1: No intersections within the same layer.}  
Curves on one layer are either Tx or Rx lines and must not intersect lest electrical cross-talk is introduced and ambiguity in sensing location arises. 
Thus, curves selected from the same layer must not intersect:
\begin{equation}
    a_i \cap a_j = \emptyset \quad \forall\, i \neq j,
    \qquad
    b_i \cap b_j = \emptyset \quad \forall\, i \neq j.
\end{equation}

\emph{C2: At most one intersection between curves from different layers.}  
Since multiple intersections between the same pair of curves lead to ambiguity in touch location, we enforce:
\begin{equation}
    \lvert a_i \cap b_j \rvert \leq 1
    \quad \forall\, a_i \in A,\; b_j \in B.
\end{equation}

\emph{C3: Bounded number of selected curves.}  
The scanning controller provides a limited number of Tx and Rx channels and the number of selected curves in each layer must not exceed these hardware-imposed limits. Let \( x_i \in \{0,1\} \) and \( y_j \in \{0,1\} \) denote binary decision variables indicating whether curve \( a_i \in A \) or \( b_j \in B \) is selected, respectively. We impose:
\begin{equation}
    \textstyle
    \sum_{i=1}^{m} x_i \leq N_{\text{drive}},
    \qquad
    \sum_{j=1}^{n} y_j \leq N_{\text{sense}}.
\end{equation}

Together, these constraints ensure that the resulting sensor layout is suitable for capacitance sensing on existing physical objects.

\begin{figure}[t]
    \centering
    \includegraphics[width=\linewidth]{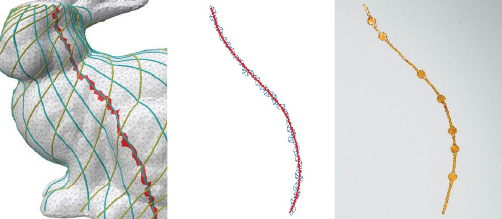}%
    \vspace{-3mm}%
    \caption{Trace fabrication. Our method unfolds each selected curve~(left),
    outlines its 2D shape and combines it with the placed electrode pads~(center),
    and sends it to a vinyl cutter for fabrication~(right).
    We use Kapton tape-covered adhesive copper tape to obtain insulated conductors in one go.}
    \label{fig:unfolding}
    \vspace{-2mm}%
\end{figure}

\subsection{Phase 2: Selecting intersections}
\label{sec:selecting_intersections}

We now optimize where electrode pads are placed, taking the set of all intersections \( I' \) from the selected curve subsets \( A' \) and \( B' \) as input.
Each intersection in \( I' \) is a candidate location for placing electrode pads.
The objective is to maximize the number of selected intersections while ensuring reliable capacitive sensing. 
This requires that electrode pads be sufficiently separated to avoid interference.
We thus impose a minimum distance constraint between any pair of selected intersections.
Let \( z_k \in \{0,1\} \) be a binary decision variable indicating whether intersection \( i_k \in I' \) is selected for electrode placement.
We formulate the objective as:
\begin{equation}
    \textstyle
    \max \sum_{i_k \in I'} z_k,
\end{equation}
subject to the constraint that any two selected intersections must be separated by at least a minimum distance \( d_{\min} \):
\begin{equation}
    \mathrm{dist}_{\mathcal{M}}(i_k, i_l) \geq d_{\min}
    \quad \forall\, i_k, i_l \in I', \; k \neq l \;\text{with}\; z_k = z_l = 1.
\end{equation}

This constraint ensures that electrode pads are separated on the surface to reduce interference and improve sensing robustness.

Phase~2 outputs a subset of intersections \( I'' \subseteq I' \) where electrode pads are rendered.
Combined with the selected curves (Phase~1), this produces the final sensor layout for wire routing and pad placement, as shown in~\autoref{fig:prototypes_intersections}.

\section{Fabrication and attachment to the 3D object}

Fig.~\ref{fig:unfolding} shows our fabrication pipeline.
Starting with the selected 3D curves on the mesh, it produces cut traces of thin Kapton-copper-Kapton laminate for attachment to the real-world object.

\subsection{Curve unfolding and outlining to planar traces}

Our method converts the 3D geodesic curves to 2D representations as described by \citet{10.11451186562} by unfolding triangle strips.
Starting from the initial triangle that contains the trace, we isometrically construct each adjacent triangle onto a 2D plane.
Since our geodesic segments are intrinsic, the unfolded traces can be reconstructed by connecting the corresponding curve nodes on the edges of each triangle. 
Once unfolded, we derive an outline of each 2D representation with 1\,mm width.
At locations where curves intersect, we union a tiny circle as a visual landmark to aid later attachment.
For the intersections selected in our optimization, our method additionally unions an electrode pad with a radius of 2\,mm, offset by 3\,mm into the routing direction.
We also union a small pad to the end of each trace for soldering wires to a digitizer.

\vspace{-2mm}%
\subsection{Electrode trace fabrication}

We fabricate all sensor traces using a three-layer Kapton-copper-Kapton sandwich structure, cut with a commodity vinyl cutter~(Cricut Maker~3).
The laminate is flexible and thin (0.05\,mm Kapton layer, 0.06\,mm copper foil). 
Unrolled, outlined, and pad-augmented traces are densely placed on a sheet for collective cutting.
The vinyl cutter has millimeter precision and supports rapid batch fabrication of all unfolded traces, cutting trace outlines in less than a minute.
We transfer each cut trace onto the object using adhesive tape.

\vspace{-2mm}%
\subsection{Attaching 2D trace outlines to the 3D object}
\label{sec:assembly}

Our approach guides precise manual attachment of the fabricated traces onto the object by projecting a textured 3D mesh onto the real-world object.
For this, our system integrates an Azure Kinect RGB-D camera to detect and track the object and an XGIMI Elfin Flip projector.
Our system segments the image using OVE6D~\cite{cai2022ove6dobjectviewpointencoding} and estimates the object's 3D rotation and translation relative to the camera based on the depth map with temporal smoothing to stabilize estimates and reduce jitter.
We project the routed curves via 3D polylines offset along surface normals and mapped to UV coordinates using barycentric interpolation for rasterization.
This step includes back-face culling to discard surfaces invisible to the projector.
We preserve spatial accuracy during projection via perspective-correct interpolation of texture coordinates.
The tracking updates in real-time, such that the 3D object can be moved and rotated during trace assembly.
Before use, we calibrate both with a checkerboard and gray-code structured-light patterns to recover the intrinsic parameters of both devices and the rigid transformation between both coordinate frames.

The projection facilitates manual attachment of the cut traces simply by following the projected lines for the bottom layer.
For attaching the traces of the second layer, we have found projection to be less helpful and instead rely on the small intersection landmarks we incorporated during unrolling and outlining (Fig.~\ref{fig:unfolding}, right) that visually guide, point-by-point, how traces are to be attached.

Finally, we solder flexible wires to trace ends and route them to a mutual-capacitance scanning controller.
Cable bundles are shielded to minimize stray capacitance and reduce sensing interference.

\section{Sensing, signal processing, and visualization}
\label{sec:sensing}

Via a soldered cable, each trace connects to a pin of an embedded mutual-capacitance digitizer IC.
Our sensing pipeline integrates a Muca board~\cite{muca_breakout}, which breaks out an FT5316DME mutual-capacitance digitizer (Micros, PL), which has been used in several prior projects for resolving touch on fabricated surfaces (e.g. \cite{teyssier2019skin,parilusyan_sensurfaces_2022}).
The chip provides 21~Tx and 12~Rx lines and reports mutual-capacitance measurements $M_{ij}$ at 10\,Hz through an SPI interface to our embedded controller.
This pin count defined the constraint values in our optimization for the objects we instrumented.
We stream these measurements to a PC via USB, where a simple processing app removes DC biases.

\subsection*{Mapping capacitive measurements to the 3D object}

The digitizer provides a matrix of capacitive measurements, commonly visualized as a 2D image.
Since our routing does not follow the traditional orthogonal grid approach, we reproject each measurement $M_{ij}$ from the mutual-capacitance matrix back to the surface location where drive line $a_i \in A$ intersects with sense line $b_j \in B$ using the known intrinsic curve layout.

\paragraph{3D visualization and contact detection.}

Mapping each intersection $i_{(a_i, b_j)} \in I$ to the corresponding face on the input triangle mesh allows us to display the registered intensity change in mutual-capacitance as a heatmap over the object surface.
This directly shows touch contacts on the 3D object with spatial correspondence to the electrode layout (Fig.~\ref{fig:teaser}, right).

Given the spatial resolution of electrodes, we perform the common interpolation of individual mutual-capacitance intensities to estimate contact areas~\cite{streli2021capcontact} when fingers or hands couple to more than one intersection.
Interpolating the intensities yields the centroid of connected components on the mesh as a single input coordinate for the purpose of detecting input events~\cite{holz2010generalized, streli2021capcontact}.
This enables a rich interaction vocabulary similar to conventional touch panels, including multi-touch, swipes, pinches, and taps---all mapped onto 3D surfaces with real-time spatial correspondence. 
For example, users can slide along a curved surface for intuitive 3D editing, or tap specific regions of a figurine to trigger animations.

\begin{absfigure}[page=7, pos=t, col=1]
    \centering
    \includegraphics[width=\columnwidth]{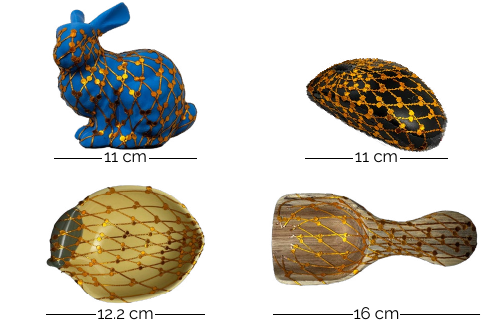}%
    \vspace{-2mm}%
    \caption{Four touch-sensitive demonstrators created by our method via object scanning, curve generation, intersection optimization, and fabrication.
    }
    \label{fig:prototypes}
\end{absfigure}

\begin{absfigure}[page=7, pos=t, col=2]
    \centering
    \includegraphics[width=\columnwidth]{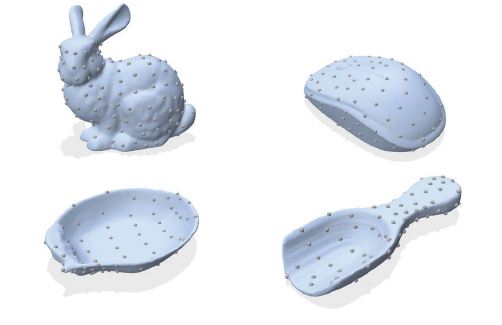}%
    \vspace{-2mm}%
    \caption{Curve-intersection locations, which later serve as sensing locations, generated by our two-phase optimization.}%
    \label{fig:prototypes_intersections}%
    \vspace{-4.5pt}%
\end{absfigure}

\begin{abstable}[page=7, pos=b, col=1]
    \centering
    \caption{Uniformity metrics of the optimized intersections.
    Lower Voronoi cell coefficients of variation (CV) indicate more uniform coverage.}
    \vspace{-2mm}%
    \label{tab:optimization_res}
    \small
    \setlength{\tabcolsep}{8pt}
    \begin{tabular}{lc}
        Object & Voronoi coefficient of variation (CV) \\
        \midrule
        Stanford bunny & 0.374 \\
        Mouse & 0.303 \\
        Plate & 0.594 \\
        Spoon & 0.451 \\
    \end{tabular}
\end{abstable}

\section{Evaluation}

\subsection{Touch sensing prototypes}

We fabricated and tested our sensing pipeline on four physical objects with varied geometry and scale (Fig.~\ref{fig:prototypes}).

\subsection{Evaluating sensor layout optimization}

\begin{figure*}[h]
    \centering
    \includegraphics[width=\linewidth]{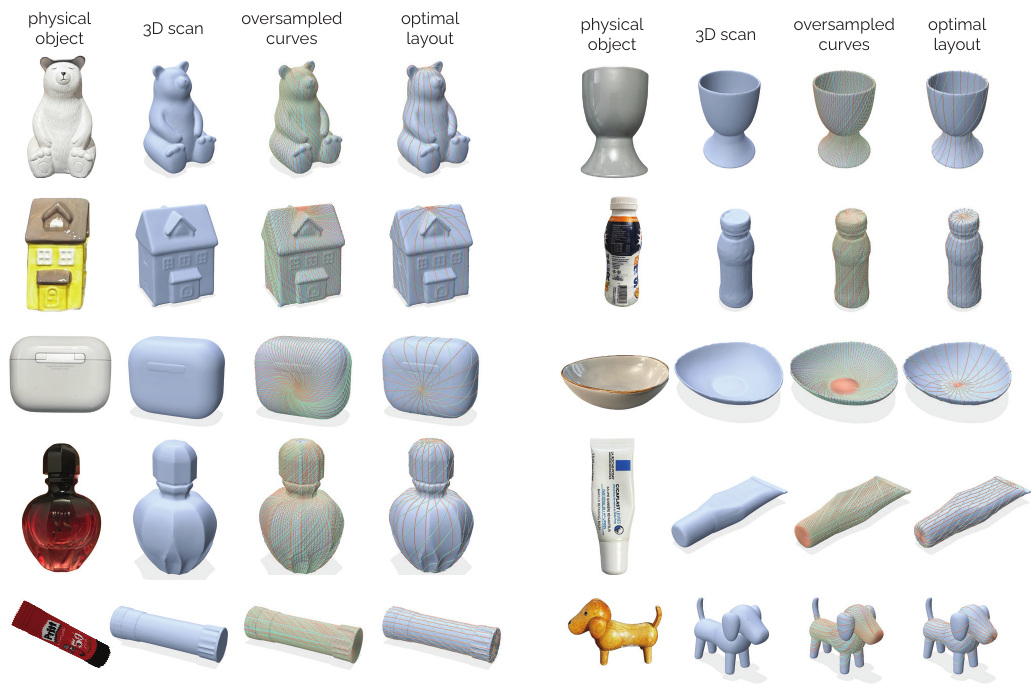}
    \caption{We apply our pipeline to ten additional existing physical objects without fabrication. For each object, we present photographs of the physical object, its 3D scan, oversampled intrinsic curves, and the corresponding optimization results.}
    \label{fig:simulation_results}
\end{figure*}

\begin{figure*}[t]
    \centering
    \includegraphics[width=\linewidth]{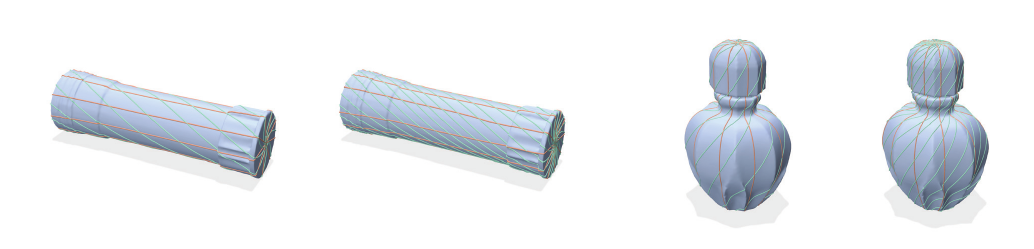}%
    \vspace{-3mm}%
    \caption{Optimization results with different numbers of curves, demonstrating support for varying resolution requirements.}
    \label{fig:prototypes2}
\end{figure*}

To evaluate spatial uniformity, we measure the coefficient of variation (CV) of Voronoi cell areas induced by the optimized intersections, where lower CV indicates more uniform coverage.
\autoref{tab:optimization_res} shows that our method produces highly uniform sensor layouts across all objects.
Fig.~\ref{fig:simulation_results} visualizes our optimization results for ten physical objects.
Fig.~\ref{fig:prototypes2} shows our optimization results under different constraints on the maximum number of curves allowed per layer, which is a limit determined by the number of pins available on the hardware sensing board.

\subsection{Capacitive sensing signal-to-noise analysis}

We measure SNR from repeated mutual-capacitance recordings across 15 touch locations per object with 10 repetitions each under varying finger orientations. 
Table~\ref{tab:snr_appendix} summarizes the measured SNR values across all objects. 
For touch sensing, SNR$\ge$15 is recommended and SNR$\ge$7 is the minimum threshold for reliable performance~\cite{davison2010techniques}.
All our SNR measurements surpassed both thresholds by a wide margin.

\begin{abstable}[page=7, pos=b, col=2]
    \centering
    \caption{Signal-to-noise ratio for touch contact.}
    \vspace{-2mm}%
    \label{tab:snr_appendix}
    \small
    \setlength{\tabcolsep}{8pt}
    \begin{tabular}{lcc}
        Object & mean SNR & SNR stdev \\
        \midrule
        Stanford bunny & 64.68 & 32.68 \\
        Mouse & 95.55 & 81.42 \\
        Plate & 64.29 & 42.72 \\
        Spoon & 52.51 & 51.04 \\
    \end{tabular}
\end{abstable}

\subsection{Localization accuracy}

We evaluate localization accuracy using 15 surface locations with 10 repeated touches per location. Touch events are detected from raw sensor signals and reconstructed to 3D positions, with no observed false positives or negatives.
Table~\ref{tab:localization_accuracy} lists the localization errors.

\subsection{Evaluating trace projection accuracy}

We evaluate projection accuracy by comparing reprojected curve intersections with manually annotated ground truth on Azure Kinect captures. Pixel distances are converted to millimeters using known object dimensions (Table~\ref{tab:intersection_distance}).

\begin{table}[h]
  \caption{3D localization accuracy (mm) across fabricated objects.}
  \vspace{-2mm}%
  \label{tab:localization_accuracy}
  \centering
  \small
    \begin{tabularx}{\linewidth}{@{}l *{7}{>{\centering\arraybackslash}X}@{}}
    Object & mean & min & max & mad & std & std$^2$ & RMSE \\
    \midrule
    Mouse & 0.76 & 0.21 & 1.24 & 0.28 & 0.32 & 0.101 & 0.82 \\
    Plate & 0.65 & 0.20 & 1.12 & 0.25 & 0.28 & 0.078 & 0.71 \\
    Stanford bunny & 1.05 & 0.20 & 1.80 & 0.38 & 0.46 & 0.214 & 1.14 \\
    Spoon & 1.60 & 0.14 & 3.39 & 1.03 & 1.10 & 1.221 & 1.95 \\
  \end{tabularx}
\end{table}

\section{Discussion}

Across our fabricated prototypes, our pipeline generates spatially consistent sensing layouts on a range of geometries without requiring access to the object interior.
Voronoi analysis shows that optimized intersections achieve uniform surface coverage across objects of varying scale and curvature.
Higher variation for the plate is due to limited sensing area and boundary effects rather than routing failure.
Overall, intrinsic curve generation and constrained optimization produce stable sensor layouts with predictable spatial resolution.
The measured SNR values show that surface-only mutual-capacitance layouts enable robust touch sensing across diverse geometries.
Mean SNR exceeds the recommended threshold for reliable detection and remains above the minimum usable level across all objects.
Higher standard deviations are due to finger orientation, contact area, and multi-intersection touches rather than sensing failures.
Overall, surface-intrinsic routing and projection-guided assembly preserve capacitive signal quality and support reliable multi-touch interaction.
Our evaluation reveals that the system achieves sub-millimeter to $\sim$2 mm localization accuracy across a range of object geometries, demonstrating that the projected electrode layouts support reliable spatial touch sensing on curved surfaces.
Projection accuracy is sufficient for practical fabrication and supports our approach.
Mean errors of a few millimeters are small relative to finger contact size and sensor spacing.
This preserves the spatial correspondence assumed by the sensing model.
Our setting is not directly comparable to \citet{palma2024capacitive}, which uses internal routing in 3D-printed objects, whereas we attach traces to existing surfaces.
Mean errors (3–4\,mm) are small relative to finger contact diameter (10\,mm) and object scale (bunny: 1.77\%, plate: 1.94\%, spoon: 2.51\%, mouse: 2.55\% of bounding-box diagonal).

\subsubsection*{Design and fabrication tradeoffs}

Our surface-only approach retrofits existing objects without modifying their volume, but the use of copper traces still occludes their appearance.
Future work could explore hydrographic printing or conductive paints and assess their sensing characteristics.
Regarding instrumentation overhead, soldering is a one-time step per object rather than per electrode.
Our Muca-based approach requires only attaching trace ends to through-holes, simplifying assembly compared to prior work such as \citet{palma2024capacitive}, which relies on internal tube routing and global surface parameterization.
Future work could further reduce overhead using alternative connectors such as snap-fit or pogo-pin solutions.

\begin{table}[h]
    \centering
    \caption{Projection accuracy onto 3D objects.}%
    \label{tab:intersection_distance}%
    \vspace{-2mm}%
    \small
    \setlength{\tabcolsep}{8pt}
    \begin{tabular}{lcc}
        Object & mean distance [mm] & distance stdev \\
        \midrule
        Stanford bunny & 3.05 & 1.56 \\
        Mouse & 3.29 & 1.98 \\
        Plate & 2.99 & 1.61 \\
        Spoon & 4.44 & 2.45 \\
    \end{tabular}%
    \vspace{3.3mm}%
\end{table}

\subsubsection*{Sensor layout generality}

Genus-0 topology covers the vast majority of practical retrofit targets, including household items, toys, figurines, and consumer products such as computer mice, bottles, and tools---making our current assumptions of genus-0 topology and a clear non-sensing base reasonable for the envisioned use cases.
While we can accommodate many practical shapes, more complex topologies will require more singularities for the computation of the discrete connections and additional evaluation for consistency. 
Narrow protrusions, such as the ears of the bunny, may receive fewer traces because only a small range of initial directions at the seed point routes curves into these regions. 
Coverage may thus be limited and reseeding or added singularities could improve it.

\subsubsection*{Sensing performance}

Beyond trace layout, capacitive sensing quality depends on aspects such as trace length, spacing, and shielding.
Our approach optimizes spatial distribution of intersections, but we do not currently account for electrical impedance or parasitic capacitance introduced by long or tightly curved traces.
Future work could integrate electrical models into the optimization to select configurations that balance spatial coverage with signal integrity.

\section{Conclusion}

We have introduced a generative computational fabrication approach for retrofitting existing 3D objects with surface-only capacitive touch sensing.
Our method formulates sensor design as a constrained geometric routing problem: it generates surface-intrinsic drive and sense lines, optimizes their intersections for spatial coverage under mutual-capacitance, hardware, and fabrication constraints, and converts the resulting layouts into planar traces for projection-guided assembly.
Our process thus preserves the object's interior and structure while adapting the multi-touch sensor design to the object's 3D geometry.
Across the four real-world objects we augmented with our method, the generated sensors achieved reliable spatial sensing, with signal-to-noise ratios well above recommended thresholds and localization errors of $\sim$2\,mm or less.
Our design studies further show that our method can produce layouts across a broader range of object shapes and sensing resolutions.

More broadly, our work shows that spatial touch sensing can be computationally designed as a retrofit for a given physical object, rather than requiring the object to be redesigned, internally instrumented, or reproduced around a predefined sensor layout.
This is particularly relevant for functional, unique, or otherwise irreplaceable objects whose geometry and interior must remain intact.
By adapting the sensing layout to the existing surface, our approach extends computational fabrication from creating interactive objects to augmenting objects that already exist.

\begin{acks}
We sincerely thank Marc Teyssier for providing us with several Muca Boards on short notice.
Zhipeng Li was partially supported by the Swiss National Science Foundation (Grant No.\ 10004941).
\end{acks}

\clearpage

\appendix

\vspace{.4cm}
\section*{\LARGE Appendix}
\vspace{.4cm}

\section{Implementation details of the optimizer}

\subsubsection*{Optimization parameters.}

We determine the minimum distance threshold \( d_{\min} \) for electrode pad placement based on the physical size of the pad.
We set \( d_{\min} \) to twice the pad diameter to avoid electrical interference between neighboring electrodes.
To approximate surface coverage in Phase~1, we uniformly sample 300~points across the surface.
We evaluated an alternative Voronoi-based sampling strategy that produced identical optimization outcomes across all 14 evaluated objects, but it was substantially slower.
Therefore, we use uniform sampling with distance-threshold rejection, which takes about 10\,s.
This provides a good trade-off between approximation accuracy and computational efficiency.

The maximum numbers of selectable curves in the two layers, \( N_{\text{drive}} \) and \( N_{\text{sense}} \), are set according to the channel limitations of the scanning controller (Section~\ref{sec:sensing}).
For the second layer, we run the full two-phase optimization independently for each sampled initial angle and select the angle that yields the best objective value across all runs.
In all fabricated prototypes, the integer linear programs returned feasible solutions. 
Should infeasibility occur, we can relax constraints by reducing electrode pad size (thereby lowering $d_{\min}$) or by reducing the number of selected curves.

\subsubsection*{Optimization solver.}
We implement both optimization phases as integer linear programs and solve them using Gurobi~\cite{gurobi}.
Runtime scales primarily with mesh resolution.
Using our setup with Gurobi's default settings, optimization for meshes of approximately 20k vertices with 21~drive and 12~sense candidate curves takes $\sim$20\,min, while meshes around 50k vertices require more than 1~hour.
In our tested range, varying pin count changed runtime only marginally compared to changing mesh size.

\section{Evaluation ablation: Effect of secondary objective}

An ablation of Eq.~\ref{eq:eq6} shows that removing this term increases the Voronoi cell area coefficient of variation from 0.3--0.5 to 0.4--0.7 across all four objects, indicating that Eq.~\ref{eq:eq6} helps promote a more uniform distribution of intersections.

%
%
\balance
\bibliographystyle{ACM-Reference-Format}
\bibliography{curved_surface-bib}

\end{document}